\documentclass{aa}
\begin{document}

   \thesaurus{08.22.3;10.03.1;  
               03.20.4;04.03.1}  

   \title{Stellar variability in low-extinction regions towards the Galactic Bulge}


   \author{T.P. Dominici, J.E. Horvath, G.A. Medina Tanco, R. Teixeira \and P. Benevides-Soares
          }

   \offprints{T.P. Dominici}
  \mail{tania@orion.iagusp.usp.br}

   \institute{Instituto Astronomico e Geof\'\i sico, Universidade de S\~ao Paulo,
              Av. Miguel St\'efano 4200, S\~ao Paulo, SP, 04301-904, Brazil\\
             }

   \date{Received  / accepted }

  \titlerunning{Stellar variability in the Galactic Bulge direction}
  \authorrunning{Dominici et al.}

   \maketitle

   \begin{abstract}

	Intensive monitoring of low-extinction windows towards the galactic 
bulge has provided in the last years valuable information for studies 
about the dynamics,
 kinematics and formation history of this part of the galaxy, mainly by 
characterizing the bulge stellar populations (Paczy\'nski, 1996).  Since 
1997, we have been 
conducting an intensive photometric-astrometric survey of the galactic 
bulge, with the monitoring 
of about 120000 stars in 12 windows uniformly distributed in galactic 
latitude 
and longitude (Blanco \& Terndrup, 1989 e Blanco, 1988) never before 
submitted to this kind 
of survey. For this purpose, we have used the IAG/USP CCD Meridian 
Circle of the Abrah\~ao de 
Moraes Observatory. The main objective of this work is the 
identification 
 and classification of
 variable objects. In this work we present the set up and development
 of the necessary tools for a project like this and the posterior 
analysis of 
our data. We briefly describe the construction of a program to organize 
and detect variables 
among the observed stars, including real time alerts (for variations
greater than 0.3 magnitudes). The preliminary 
analysis after 
the processing of 76 nights
 of observation yielded 479 variable stars, from which 96.7 \% of them 
are new. 
We discuss the preliminary classification of this variables, based on: 
a) the observed 
amplitude of variation; b) the shape of light curve; c) the expected 
variable classes 
among our data and d)  the calculated periods, whenever possible.
 Finally, we discuss the future perspectives for the 
project and for the applications and analysis of the 
discovered variable stars.

   \end{abstract}

%

\section{Introduction}

The picture that we formed of the Galaxy suffered many
modifications since the first analysis of the count of stars in the 
eighteenth century. After
the discovery of the existence of large amount of gas an dust, which 
causes light extinction and
makes the observations of the center region of the Galaxy very difficult
in several bands, the 
conclusion was that the Galaxy is of 
a spiral-type, with arms composed by metal rich young
stars and a halo and the central region (bulge), constituted by metal
poor old stars.

It has been later discovered that the irregular distribution of dust and 
gas presents some "holes" 
where the extinction is considerably lower. The best known of these 
so-called 
{\it windows} has been found by Baade (1951) and extensively studied 
since then. 
Generally speaking, these low-extinction windows offer a unique 
opportunity 
to obtain information about the bulge in the optical range and compare
this part of the Galaxy with the halo and disk (and
with other galaxies). Great progress have been obtained
by the observation of these regions and our picture of
the Galaxy changed. For example, with the advent of direct measures
of metal abundances of the K giants in the Baade window (Rich, 1988), it was
possible to conclude that the Galactic bulge is 
not a metal-poor structure: its  
mean metallicity is twice the solar neighborhood value. 
This means that the halo
and the bulge are quite distinct structures. 
Today we know that in the galactic bulge 
reside stars showing a large range of metallicities.

However, many questions still remain open. 
The knowledge of the age of each galactic 
component
is one of these and it is important to decide among the several 
formation 
scenarios: would  the bulge be older than the halo, and if so by how 
much? 
How these structures formed? Many authors have worked 
on these subjects. Terndrup (1988)
estimated the bulge age using the color-color diagrams 
of four low-extinction
windows, analyzing the position of the main sequence strip. He concluded 
that the bulge is between 11 and 14 Gyr old. In another work, Lee (1992) 
calculated the relative age of the RR Lyrae in the halo and in the bulge and 
found 
that bulge stars would be older than the halo RR Lyrae by 1.3 Gyr.
The discovery of very old stars in the bulge does not necessarily mean 
that the bulge is the first structure to be formed, since the possibility of a 
merger population exists.

Recently, several works have shown the presence of a galactic bar in
the bulge. Stanek et al. (1994), using the stars of the red clump, found 
an asymmetry in galactic longitude. Similar results were obtained  with
the hydrogen 21 cm line (Liszt \& Burton, 1980) and in 2.4 $\mu$m
(Blitz \& Spergel, 1991). Whitelock and Catchpole (1992) 
detected a longitudinal
asymmetry in the distribution of the Mira variables in the bulge. 
The actual extension and inclination of the bar are problems yet to be
solved.

Many works to help to solve this and other problems related
to the galactic bulge are possible 
provided adequate data is available. Besides their intrinsic 
interest as representatives of specific evolutionary phases, variable
stars are one of the most valuable ones since they can be used as 
distance indicators, tracers of metallicity gradients and several other 
applications, being a very useful tool for galactic structure investigation.

Some massive monitoring programmes, as examples
MACHO and OGLE groups, have been recently monitoring the bulge in 
search of microlensing events (see Paczy\'nski (1996) for a review) with 
well-known success. The secondary
products, complete catalogs of the monitored regions and the discovery of new 
variable 
stars, proved to be as interesting as the microlensing events 
themselves. However, most of
these projects are indeed monitoring a limited area of the galactic 
bulge (in their first phases) in order to maximize the microlensing 
detection probability. Therefore, 
a wider coverage of stellar fields for other relevant regions is not 
attempted.

With these limitation in mind, we have developed and conducted 
a small monitoring project of 12 selected low-extinction windows 
from April 1997 until August 1998 (see Table 1). The windows were taken 
from the works of Blanco \& Terndrup 
(1989) and Blanco (1988). In two successive campaigns  we have used 
the recently refurbished Meridian 
Circle of the Abrah\~ao de Moraes Observatory (operated by the 
IAG/USP) to observe these fields (see below).
The main objective of the project was the discovery and classification 
of variable stars (in principle, for variations
greater than 0.3 magnitudes), with the construction of a database
that can be used by the astronomy community for several researches, as 
discussed above.
The final goal will be to have an on-line, real time 
processing data to stimulate the study of potentially interesting events 
by other observing facilities. The first scientific 
result of the project has been a high-quality extension of the Tycho 
catalog for the observed regions (Dominici et al., 1999, hereafter Paper 1), 
a necessary step to perform the data reduction.
We should remark that this project was not appropriate for the search of 
microlensing events since the
meridian circle allows the observation stars up to $V$ = 16 
approximately and the probability to detect this kind of effect is low.

The use of the meridian circle for 
photometry purposes was attempted previously by Henden 
\& Stone 
(1998), who published a catalog of 1602 variable stars, 
of which only 85 were found in the literature. The observations
which originated this work were performed by FASTT, which is
very similar to our instrumental setup.

Many types of variables were expected to show in our project.
Since we were working with a small refractor instrument, only the 
brighter stars
of the bulge, like the Miras, could be observed. Stars that are present
in the disc, like classical cepheids are expected and the intermediate 
population (the transition between disc and bulge and 
between halo and bulge),
like RR Lyrae and W Virginis, should be also present in the sample. 
Binary systems can be found
in all galactic regions along the sight of view, having a variety of 
magnitudes, 
periodicities and stellar components. Cataclysmic variables should be 
relatively rare but fully detectable with our instrumentation (Dominici et al., 
1998a).

We shall describe in the next Section the selected fields. Section 
3 is devoted to a presentation of the instrumental facilities and data 
reduction method. The observational program and the differential 
photometry with the Meridian Circle
are described in the Section 4. The development and tests of the program to 
organize and search for variable objects ({\it Class32}) is 
discussed in Section 5. The variable stars found and their
classification and analysis methods are detailed in Section 6. 
Conclusions and 
future perspectives are presented in Section 7. 
The light curves examples in Appendix A, 
the catalogue of variable stars in Appendix B\footnote{Tables B.1
 to B.12 are only
available in electronic form at the CDS via anonymous ftp to
cdsarc.u-strasbg.fr (130.79.128.5) or via
 http://cdsweb.u-strasbg.fr/Abstract.html} and
a brief description of the effects of
the chromatic aberration in the meridian circle in Appendix C 
close the present paper. 


\section{Selection of the fields}

With the aim of studying M giants in the galactic bulge, Blanco \& 
Terndrup (1989) and  Blanco (1988)
collected in the literature low-extinction windows evenly distributed in  
galactic latitude and longitude and selected new regions with analysis 
of old photographic plates. The distribution of the resulting 31 windows, that
we adopted for our observational programme, can be seen in Figure 1. 
Since we needed to develop the necessary tools and
analysis methodology to start the project, we decided to
use only 12 windows for the first stage of the work.
Table 1 summarizes the list of these low-extinction windows, as 
displayed in Figure 1.

\bigskip

Figure 1: Position of the center of the 31 low-extinction windows 
collected or discovered by Blanco \& Terndrup (1989) and  
Blanco (1988). The hollow symbols 
indicate the 12 windows that we monitored in the 
first stage of the project.


\bigskip

The great advantage of this sample of windows is that they cover
 almost all bulge area. The monitoring programmes with the aim to
discover microlensing events (MACHO and OGLE I, for examples) 
are concentrated in areas near the
Baade window, basically in the interval $-6^{o} < l < 6^{o}$ and
$-6^{o} < b < -2^{o}$ (Stanek et al., 1996, Bennett et al., 1994).
In this sense, our shallower but spatially extended search can be 
considered as complementary of the massive databases of the central 
regions. 

\begin{table}[tb]
\caption{Coordinates of the 12 low-extinction windows that we choose
for monitoring in the search for variable stars.}
\label{tabrmm}
\begin{center}
\begin{tabular}{@{}lcrcrr@{}}
\hline
\rule{0pt}{1.2em}Window&  \# & $\alpha$  (J2000) &$\delta$ (J2000)  &  b 
($^{o}$)&l ($^{o}$)\\
\hline
BE  &    5   &    18 10 17    &    -31 45 49    &   -6.02  & 0.20 \\  
BG  &    7   &    18 18 06    &    -32 51 24    &   -7.99  & 0.00  \\
BJ  &    10  &    l8 40 49    &    -34 49 48    &   -13.10 & 0.27  \\
LA  &    11  &    16 56 57    &    -52 44 53    &   -6.02  & -24.77 \\
LB  &    12  &    17 18 27    &    -48 05 40    &   -6.01  & -18.96 \\
LC  &    13  &    17 29 22    &    -45 20 05    &   -6.03  & -15.61 \\
LD  &    14  &    17 44 59    &    -40 41 37    &   -5.99  & -10.14 \\
LI  &    19  &    18 04 48    &    -33 47 46    &   -5.96  & -2.11 \\
LR  &    27  &    18 23 32    &    -26 10 16    &   -5.98  & +6.54\\
LT  &    29  &    18 34 09    &    -21 23 35    &   -6.04  & +9.87 \\
LU  &    30  &    18 51 26    &    -13 18 41    &   -6.03  & +21.04 \\
LV  &    31  &    18 57 37    &    -10 23 35    &   -6.08  & +24.35 \\
\hline
\end{tabular}
\end{center}
\end{table}


\section{Instrumental facilities and data 
reduction}

The Askania-Zeiss Abrah\~ao de Moraes meridian circle, operated at the 
IAG/USP Valinhos Observatory ($\phi$ = $46^{o}$ 58' 03'', $\lambda$ =
$-23^{o}$ 00' 06''), 
is a 0.19 m refractor instrument with a focal distance of 2.6 m. 
A CCD detector Thomson 7895A with  
$512 \; \times \, 512$ pixel matrix and a pixel scale of $1.5"/pixel$ 
was used for the imaging (Paper 1, Viateau et al., 1999).  
The observations are performed in a drift-scanning mode.
So the integration time, for a declination $\delta$, 
is given by

$$ t_{int} \, = \, 51 \, \sec \delta \, \, s    \, \; \;       \eqno(1)$$

The observed field has a width of $13'$ in declination by an arbitrary 
interval in right ascension (some minutes to several hours). It is possible 
to obtain images of several thousands of stars per night, up to 
magnitudes as faint 
as $m_{Val} \, = \, 16.5$, depending on the transit time as reflected 
by eq.(1) (we will use "$m_{Val}$" to designate the
 magnitudes obtained with the Valinhos system ($V_{V}$ filter) and
"$V$" for magnitudes in the standard Jonhson system ($V_{J}$ filter), 
see below).

The optimal magnitude interval for the 
observations is  $9 \, < \, m_{Val} \, < \, 14$, while
the typical accuracy in the positions and magnitudes for a 
single measurement of a given night is shown in Figure 1 of 
Paper 1.

The filter we have used in 
this and other observational programmes is somewhat wider than the 
standard Johnson filter $V_{J}$; allowing a larger coverage towards the infrared 
band in order to maximize the number of objects by taking advantage of the 
better quantum efficiency of the CCD in that region. Figure 2 of Paper 1
 shows the response of 
the Valinhos filter $V_{Val}$ together with the standard Johnson filter 
$V_{J}$. The correlation between the filters, the method used for our 
programme and the limitations are
described in Paper 1. In the present work we have used the differences 
of magnitudes of the stars with respect to a standard 
reference set, as described in the next Section.

The employed data reduction method requires a first
step, where the sky background is subtracted by a linear polynomial fitted to 
each pixel column. Objects are identified when 3 consecutive pixels with a 
$2\sigma$ confidence level are detected, where 
$\sigma$ is the standard deviation of the mean count rate in each 
column. A two-dimensional Gaussian surface is fitted to the flux distribution of 
the objects, to 
obtain the $x$ and $y$ coordinates of the centroid, the flux and 
respective errors (Paper 1, Viateau et al. 1999).
In the following step, the celestial positions and magnitudes are 
calculated by solving a system (equations 3-5 in Paper 1) with respect
to reference stars (Section 4).
For our variability analysis we take these results, i.e.,
those obtained in the classical way, night-by-night.

After this process, the system is again solved in an 
iterative process, now for all stars detected in the field, by
a global reduction, using the field overlap constraint among all observation nights 
(Eichhorn, 1960, Benevides-Soares \& Teixeira 1992, and Teixeira et al. 1992).
 At each step of iteration, the system is 
solved by least squares (Benevides-Soares \& Teixeira; Teixeira et al. 
1992). The process
converges in a few iterations (typically less than 10 steps). This
method is applied to obtain the best values to the positions, as can
be seen in Appendix B.


\section{Observational programme and the photometry with the Meridian Circle}

The project was idealized to observe the selected windows in all nights 
of good weather, whenever the bulge was visible. Two campaigns were 
concluded: the first are between April and September, 1997 and the second are 
between April and 
August, 1998. The average size of the (irregular) fields was 3 to 6$^{m}$ in 
right 
ascension and 13' in declination (see Table 2). 

Since the two refracting objective presents considerable secondary
chromatic aberration, the spectral band should be centered around
the focal minimum at $\lambda$ = 498 mm.
The CCD response peak at 700 mm and a filter was added to create
a final spectral band between 500 and 900 mm. The spectral
limitations prevent the use of a second color band filter.

\begin{table}[tb]
\caption{Initial and final observation sidereal time for the 12 windows 
chosen for the Valinhos monitoring programme.}
\label{tabrmm}
\begin{center}
\begin{tabular}{@{}lcc@{}}
\hline
\rule{0pt}{1.2em}Window& Initial TS($^{h}$ $^{m}$)& Final TS($^{h}$ 
$^{m}$)\\ 
\hline
LA &	16 54&	17 00\\
LB &	17 15&	17 21\\
LC&	17 26&	17 32\\
LD&	17 42&	17 47\\
LI&	18 03&	18 06\\
BE&	18 08&	18 13\\
BG&	18 16&	18 20\\
LR&	18 22&	18 26\\
LT&	18 32&	18 36\\
BJ&	18 39&	18 44\\
LU&	18 48&	18 53\\
LV&	18 55&	19 00\\
\hline
\end{tabular}
\end{center}
\end{table}

Initially, we had to face the problem that the total number 
of reference stars of the usual (astrometric) catalogues
inside the defined fields is small. This posed a serious difficulty for 
data reduction and the search for variable objects. As a first 
step towards a comprehensive study of the windows, we have attempted to 
construct dense (secondary) reference catalogues, based on
Tycho catalog (ESA, 1997), intended to be of general use. 
Methods and final catalogs are given in Paper 1. For their 
construction we needed to make "long" exposures of about 1 hour 
in right ascension to include as 
many Tycho stars as possible in each field.
During the first 1997 campaign we performed 5-6 "long" 
frames for each window. Tycho catalogue was extended 
for stars up to $m_{Val} = 13$, limited to the non-variable stars. 
A summary of the results, with the mean precision and the number of secondary
reference stars obtained for each window is presented in Table 3.

\begin{table}[tb]
\caption{Average position and magnitude  precisions for secondary reference 
stars 
in the low-extinction windows (mean epoch JD 2450612). 
Last column indicates the number of reference stars in the final 
catalogs.}
\label{tabrmm}
\begin{center}
\begin{tabular}{@{}lcrccc@{}}
\hline
\rule{0pt}{1.2em}Window& $\sigma_\alpha(^{s})$& $\sigma_\delta('')$& 
$\sigma_{m_{Val}}$& $\sigma_{w}$& N\\
\hline
BE     &	0.0017	&	0.015	&	0.028 &		0.0043 &  60\\
BG     &	0.0021	&	0.014	&	0.026 &		0.0040 &  41\\
BJ     &	0.0021	&	0.013	&	0.027 &		0.0041 &  43\\
LA     &	0.0019	&	0.011	&	0.035 &		0.0038 &  55\\
LB     &	0.0019	&	0.011	&	0.033 &		0.0039 &  36\\
LC     &	0.0014	&	0.011	&	0.029 &		0.0035 &  61\\
LD     &	0.0015	&	0.011	&	0.027 &		0.0036 &  47\\
LI     &	0.0010	&	0.011	&	0.031 &		0.0038 &  20\\
LR     &	0.0025	&	0.014	&	0.031 &		0.0048 &  37\\
LT     &	0.0021	&	0.014	&	0.028 &		0.0049 &  32\\
LU     &	0.0017	&	0.013	&	0.028 &		0.0045 &  31\\
LV     &	0.0015	&	0.016	&	0.034 &		0.0048 &  44\\
\hline
\end{tabular}
\end{center}
\end{table}

At the end of the two campaigns, we had collected a total of 76 nights with best 
quality images (average of 35 observations per window).

It is useful to remark that some standard 
steps in the frame treatment for photometry 
(flatfield, bias) are not necessary nor possible in our case: in drift 
scanning observation mode each "pixel" actually 
corresponds to a combination of 512 pixels, and then, even if the 
response of each elementary pixel is known to be 
non-uniform, the final result is the column-wise sum 
over the CCD 512 pixels width (the columns are parallel to the
image motion). The summed line so obtained displayed an
essentially constant response during several flatfield measurement
tests, so that the field correction can be dropped.

Given that the integration time is limited by the time of stellar 
transit, the telescope is
small and the site quality is not very high, 
we could only observe stars up to $m_{Val} < 16.5$, approximately. 
Therefore, we can detect only to brightest bulge stars and 
most of the monitored stars are actually foreground objects.
The completeness of the sample for BE window is shown in Figure 2.
The maximum of the histograms for each window can be seen in table 4.

\bigskip

Figure 2: Completeness of our sample in the BE window.


\bigskip

\begin{table}[tb]
\caption{Completeness of the sample for the monitored 12 low extinction 
windows. The $Mag_{max}$ column indicates the faintest objects detected for
the region and the $Hist_{max}$ column gives the magnitude corresponding
 to the largest frequency}
\label{tabrmm}
\begin{center}
\begin{tabular}{@{}lcc@{}}
\hline
\rule{0pt}{1.2em}Window& $ Mag_{max} $& $Hist_{max}$\\
\hline
BE     &	18.3	&	16.0	\\
BG     &	18.2	&	16.0	\\
BJ     &	18.1	&	16.0	\\
LA     &	18.4	&	16.4	\\
LB     &	18.3	&	16.2	\\
LC     &	18.2	&	16.2	\\
LD     &	18.2	&	16.0	\\
LI     &	18.2	&	15.8	\\
LR     &	18.1	&	16.0	\\
LT     &	18.1	&	15.6	\\
LU     &	18.1	&	16.0	\\
LV     &	18.0	&	16.0	\\
\hline
\end{tabular}
\end{center}
\end{table}

With the aim of removing systematic errors, we 
worked with the difference of magnitudes with respect to the mean 
magnitude of the reference stars
in the field which compose the secondary references catalogues presented 
in Paper 1 (or {\it differential magnitudes}). 
The idea was to remove systematic errors due to observational 
conditions that affect much in the same way the reference stars 
as well as all other objects in the field. It should be stressed that,
in general, 
this removal procedure is more efficient when the comparison 
reference stars have magnitudes and colors comparable to 
the field stars and, therefore, in our case the use of 
differential magnitudes was not completely efficient, according
to the tests. However, since 
we monitored more than 120000 objects every night a less general 
treatment was not feasible. 

Some general criteria were adopted to evaluate if the frame and 
final reduction had
sufficient quality so as not to cause systematic effects in the analysis 
and search for variable stars in the databases, like spurious alarms of
variability.

The first of these criteria is the comparison of 
the quadratic sum of the differences between the night and 
the mean values.
Whenever these differences were $\geq \,  0.1$ in magnitude, the frame 
was eliminated. 
Analogously, the results in right ascension and declination 
were inspected to check whether the behavior of the
residues (the difference between the night and the mean values) of the
magnitude as a function
of the sidereal time was abnormal. Figure 3 shows this behavior for LI 
window, note that problematic nights and the identification of the night effects 
are clearly visible in the plot and caused the elimination of unsuitable 
data.

\bigskip

Figure 3: Residues of the magnitude in function of the sidereal time. 
Each row represents a single  night of observation. Several effects that 
provoke the frame elimination are indicated in the figure. It is possible
to note the increase of the number of monitored objects in the 1998 
campaign.


\bigskip


\section{Searching for variable stars}

To organize and analyze such amount of data is not a 
trivial work, specially if the final goal is 
to investigate the presence of variable objects 
in the monitored regions in real time.

A program called {\it Class32} was developed by us 
to handle the huge volume of data.
Basically the 
program constructs a database for each investigated window, where
a main table contains mean informations about all objects
already detected, and an individual record for each object as well. 
In the main table we record\\
\\
\noindent
\_ A sequential numerical identification;\\
\_ The mean value of the right ascension;\\
\_ The mean value of the declination;\\
\_ The standard deviations corresponding to right ascension and 
declination;\\
\_ The mean differential magnitude;\\
\_ Standard deviation of the mean differential magnitude;\\
\_ The number of times that the object was detected;\\
\_ A Flag that indicates the confidence of the identification;\\
\_ Gregorian date of the last observation;\\
\_ A second flag (hereafter, AlarmFlag) that warns if the object
suffered light variations and in how many days.
\\

The individual table for each object contains (for each night):\\
\\
\noindent
\_ Gregorian observation date;\\
\_ The calculated right ascension;\\
\_ The calculated declination;\\
\_ Differential magnitude with respect to the reference set;\\
\_ The magnitude error;\\
\_ The AlarmFlag;\\
\_ Apparent magnitude;\\
\_ Julian Date of the observation.
\\

Now, we will describe in more detail the contents of 
the tables. The
input of this program are the positions values (right ascension
in hours, declination in degrees), differential magnitude,
magnitude, the respective errors and the Julian Date.
For a given night, the input positions 
of the detected stars are compared with the mean values available in 
the main table. The first criterion for identification is that
the centroid falls within the same pixel.

The second criterion, useful for the 
confirmation of the identification is

$$ \bar{x} + d\bar{x} > x - dx \, \eqno(2)$$

$$ \bar{x} - d\bar{x} > x + dx \, \eqno(3)$$
for the right ascension and declination, where $\bar{x}$ and $d\bar{x}$
indicate the mean value and its error and $x$ and $dx$ are
the night values.
The criterion of Eqs.(2) and (3) verifies if the last measurement, 
allowing for the error, is inside the region 
delimited by the mean value, calculated from 
the previous nights, plus one standard deviation.

After the identification, the program makes 
a series of hierarchical tests to verify if the object
presents light variations. The first test is 
somewhat arbitrary and  connected with
the variations that we consider easily measurable by our project

$$  \left\vert{\bar{m} - m}\right\vert > 0.3 \, \eqno(4) $$
where $\bar{m}$ is the mean differential magnitude.

The next test, which is applied only if the star satisfies 
Eq.(4), is more refined and
compares the night measurement with the mean value, 
taking into account the
standard deviation of the mean differential magnitude ($d\bar{m}$)

$$ m > \bar{m} + nd\bar{m} \, \eqno(5)$$

$$ m < \bar{m} - nd\bar{m} \, \eqno(6)$$
where $n = 3$ was adopted after extensive testing of the method.

The last test, which is applied only if the star 
passed through the tests of Eqs.(5) and (6), 
takes into account the "instantaneous" error of that particular night 
and reads

$$  \left\vert{m - \bar{m}}\right\vert < \left\vert{dm}\right\vert  + 
 \left\vert{d\bar{m}}\right\vert \, \eqno(7)$$ 

Even if the star satisfies the criterion of Eq.(4) only, the program 
registers an
alarm (the AlarmFlag). Our experience shows that
stars passing only in the first criteria are those near the detection 
limit 
and have large photometric errors. Most of the stars that
show a real (and reliable) variable behavior have passed 
through all criteria.

As a test of the performance of the {\it Class32} program 
and with the aim of verifying our actual observational capabilities, 
we have observed the microlensing event 97-BLG-56, detected by the 
MACHO group, as a target-of-opportunity. Since the star was known
to vary and other groups were monitoring it, a comparison and
evaluation of the actual behavior of the devised variability
criteria was possible in a quite accurate form.
We obtained 16 good quality frames of the event. The reduction was
made of the standard way but, since the field does not belong to 
any of the
monitored fields, we had to calculate the differential magnitude with
respect to the available Tycho stars in the same 
field, subtracting the stars brighter than
$m_{Val}$ = 8.5 and those having a Tycho quality index worse 
than 5 (see Paper 1).

The identification of the ongoing event 
in the database was sucessfull and the
program gave the expected alarm in 11 consecutives nights. The light 
curve of the event can be seen in Figure 4, where our data
was superimposed to the OGLE I filter measurements (Udalski et al., 1997). 
Our data, and the subsequent analysis of
the 97-BLG-56 will be subject of an future paper.

\bigskip

Figure 4: Comparison between our observations
(in $V_{Val}$ filter) and OGLE data (I filter, 
taken from Udalski et al., 1997). The axes indicate
the difference between the stationary value of the 
magnitude (m) and the night measure during the amplification
(m(t)), versus the Julian Date. 


\bigskip
 
Real time processing of the data, though 
being the ultimate goal of the programme, was not 
indeed possible in
this first stage of the project. The {\it Class32} 
is still under development and presently the time 
it takes for a full processing of one night 
is greater than expected. We will discuss
the future modifications in our analysis methods in the last Section.


\section{Variable stars analysis: the minimum entropy method}

Because of poor weather conditions and unexpected instrumental 
problems the 1998 campaign was finished slightly 
earlier than planned in August 1998. The final number
of observations including both campaigns was less than 
originally expected: a total 76 good quality nights with an 
average of 35 observations per window. Even if the data is 
good enough for the discovery of variables, this has
limited our analysis of the variables found (see below).

When we finished all the analysis with the {\it Class32} 
program, we had finally obtained data from 
124976 objects in the 12 databases. The AlarmFlag
of each object was checked and the stars
that presented a significant variation (sufficient
number of observations, reasonable photometric
errors, large magnitude variations, etc.) 
selected. The total number of objects per window
and number of variables found are in Table 5. 

\begin{table}[tb]
\caption{The final number of objects and variable stars
found in each window.}
\label{tabrmm}
\begin{center}
\begin{tabular}{@{}lcc@{}}
\hline
\rule{0pt}{1.2em}Window& $n^{o}$ of stars& $n^{o}$ of variables\\ 
\hline
BE&	14750&	74\\
BG&	10114&	54\\
BJ&	5941&	15\\
LA&	12952&	79\\
LB&	10535&	33\\
LC&	12466&	32\\
LD&	10628&	35\\
LI&	9006&	32\\
LR&	11820&	56\\
LT&	8454&	23\\
LU&	8630&	20\\
LV&	9680&	26\\
Total& 124976&   479\\
\hline
\end{tabular}
\end{center}
\end{table}

Among the variables we have found (see the complete listings 
in Appendix B), a  major subset  shows hints of
periodic variations. However, if we do not have 
clear evidence to say
that the star is periodic, its classification is highly
compromised. As stressed before, the main problem is 
that we do not have as yet chromatic/spectral 
information about the objects. Since the variable classes are defined
mainly by the spectral characteristics (especially in the case of
aperiodic variable stars), the study of the individual 
stars given in the Appendix B is a natural and necessary 
next step. It should be remarked that 
the temporal covering of our data is not (generally 
speaking) broad enough to
affirm whether a given star that does not show periodic variations 
is actually aperiodic.

In the case of the stars with hints of periodic light curves it is
important to estimate the period (or periods) for a tentative 
classification. Several algorithms to perform period
calculation exist, the Fourier method is the most popular. 
However, as is well-known the
Fourier method works with equally-spaced points from a temporal series,
which is never the case for astronomical measures. Many authors 
(see, for example, Cuypers 1987) have 
developed a modified Fourier analysis for non-equally-spaced 
points. These works have generally resulted in computational 
time-consuming 
algorithms. In the our analysis
we have adopted a new method (the minimum entrophy method), developed by
Cincotta et al. (1995).

This method is based on the minimization of the information entropy for
the true period(s). The rigorous mathematical formulation
can be found in Cincotta et al. (1996), for example. We will
quickly describe the method.

Consider a temporal
series ($u(t_{i})$) and the set of periods to test $p_{j}$ $(j =
1,...,n)$.
According to Cincotta et al. (1995) we have calculated the phase of the 
temporal 
series and created an unitary 
(normalized) plane $(\phi,u)$,
where $\phi$ is the phase.
For each trial period $p_{j}$, the light curve has been constructed and 
distributed
in  $(\phi,u)$ plane, that is divided in an arbitrary number ($N$) of 
partitions.
The next step was to calculate the probability ($\mu_{i}$) for a point 
to fall
in one partition, dividing the number of points in each partition by the
total number of points in the temporal series. Finally we have computed 
the
entropy as

$$ S = -\sum_{i=1}^x\mu_{i}ln(\mu_{i}) \, \eqno(8)$$

If the trial period is not the real one, the light curve points will be
uniformly distributed in the plane $(\phi,u)$ and the entropy 
will be maximum. On the other hand, if the trial period 
matches is the actual period, the points will  be limited
in to small region of the plane and we will have a minimum of the 
entropy. 
Therefore, we expect that minima in the $S$ vs. $p_{j}$ diagram 
correspond 
to the actual period(s).

The method works very well when the light curve is sufficiently sampled 
(empirically, $\geq$ 50 points for the period determination 
to be strictly independent of the partition
number (N)). This is not quite the case of our databases, 
although the actual number of points is not too low 
in some cases. Thus, 
the application of the method remains 
feasible, but noisier $S$ vs. $p_{j}$ diagrams are obtained. 
On the other hand, the computing processing time is less 
time-consuming than the modified Fourier analysis and, in principle, we 
believed 
suitable for our extensive databases
(see discussion below). Another well-known 
problem is the presence of
the harmonics, with can cause some confusion when we have few points.

Simulations and comparisons between minimum entropy and modified 
Fourier analysis were performed by Cincotta et al. (1995). Their tests 
showed that the minimum
entropy method is more efficient for the resolution of multiple periods
than the Fourier analysis. However, the accuracy of the estimated periods
are not known yet and we intend to collaborate with the development of the method
with its massive application to your databases.
A comparison between several methods to determine periods present in 
our data will be the subject of a future study.

For simplicity, we consider that
the stars have only one period. We shall present some examples from our
 databases to see how this 
works and evaluate the results. As we say before, the method is independent of
the partition number when the series is well sampled, this isn't our case and,
for each star, the calculations
were done for several configurations of parameters,
 varying the number of partition (N) and trial period interval. We considered a period
as found when the minimum persisted for all tried configuration 
(the differences between minima in each configuration are in the less significative digits).

 The star LI471 is a W Virginis
already known before and belonging to GCVS (General Catalogue of 
Variable Stars,
Khopolov et al., 1988), with an attributed 11.49 days period. The light
curve of the star (with our data) 
can be seen in Figure 6.
Figure 5 shows the $S$ vs. $p_{j}$ diagram for that 
star which is noisy as expected. 
Nevertheless, the minimum we have obtained 
by applying the minimum entropy method is 11.50 days, consistent with 
the 
catalogued
value. This example shows that, in spite of the noise, accurate 
determinations are possible in an efficient form.

\bigskip

Figure 5: $S$ vs. $p_{j}$ diagram for the star LI471, a known
W Virginis.


Figure 6: Light curve of the star LI471.


\bigskip

Besides the problem of the scarcity of the points, our photometric 
errors 
are quite large, which in turn 
provokes more noise in the diagrams. As a result of the analysis we have 
checked 
that the determination of small periods is 
easier and more reliable than the determination of large periods (like 
Miras). 
This fact is closely related to the definition of the phase 
(see Cincotta et al. 1995) which is inversely proportional to the period.
To give an example, we discuss the star LA1552, not known before and 
preliminary classified as a Mira, which
reflects this difficulty very well (the light curve
can be seen is Figure 8). Figure 7 shows the $S$ vs. $p_{j}$ 
diagram. We estimate the period in 255.4 days, but in this case the
method is 
more dependent on the number the partitions.

\bigskip

Figure 7: $S$ vs. $p_{j}$ diagram for the star LA1552, classified
as a Mira. The estimated period is 255.4 days.


Figure 8: Light curve of the star LA1552.


\bigskip

As a further example of the capability of the method 
to find periods smaller
than 1 day, we show the case of BJ878, a RR Lyrae star already known 
and belonging to
GCVS (Figure 10). The catalogued period is 0.37 days and our best 
estimation was
0.33 days. Figure 9 shows the $S$ vs. $p_{j}$ diagram, with
a well resolved minimum. In fact, it is not impossible that the 
catalogued period has to be corrected if further studies confirm 
the present value.

\bigskip

Figure 9: $S$ vs. $p_{j}$ diagram for the star BJ878, a
RR Lyrae. The estimated period was 0.33 days.


Figure 10: Light curve of the star BJ878


\bigskip

Even in cases where periodicity is suggested by the data, we 
have not been able to determine period for all those candidate stars. 
Some of them have so
few points that it was impossible to draw a reliable conclusion. These 
stars are indicated with "NC" (not calculated)
in the nineth column in the catalogue
presented in Appendix B. The stars that we believe
to be periodic, but for which a period could not be reliably found, 
are indicated with "NF" (not
found) in the same column.

To perform a preliminary variable star classification, we 
have based our judgement on:\\
\\
\noindent
\_ The available periods as calculated with the minimum
entropy method;\\
\_ The observed amplitude of variations;\\
\_ The shape of light curves, which have been compared 
to the observed for variables already existing in the GCVS;\\
\_ The variable classes that we expected {\it a priori} among the
monitored stars, as described in the Introduction.
\\

All stars positions were compared with the ones of the variable 
objects belonging
to the GCVS and  NSV (New Suspected Variables, Kukarkin et al., 1982) 
catalogues  and 
to those of the SIMBAD database
using WWW interface. Among the 479 stars that showed 
significant light variations found in our database,
only 16 were already present in other catalogues (including 
the IRAS catalog (Beichman et al., 1988), the work of Hazen (1996) on 
the variables in NGC 6558 (in the center of BE window), 
the CCDM (Catalog of Double and Multiple stars) by Dommange (1983),
besides the quoted GCVS and NSV catalogues). Therefore, 
 96.7 \%  of the variables
in the new databases were unknown until now.

Considering our limitations, we restricted the classification 
of our variables to four broad classes {\it defined} as\\
\\
\noindent
\_ Mira-type variables: stars with period greater than 80 days, 
amplitude of the variation greater than 2.5 magnitudes (approximately), 
and
light curves alike BG1159 (Figure 11, Appendix A), a Mira 
already present in GCVS;\\
\_ Semi-regular-type variables: stars with periods greater
than 80 days and light curves alike the Miras, but with 
amplitude variation smaller than 2.5 magnitudes;\\
\_ Cepheid-type variables: stars with periods between 1 and 50 days,
(approximately), and mean amplitude variations of 0.9 magnitudes.
The differentiation between classical cepheids, that populate the
galactic discs, and W Virginis (or type II cepheids), that are
present among the older population (halo, bulge), is possible only 
with the knowledge of
morphological differences in their light curves. Since we are not 
sensitive to these
details, we have classified these stars as"cepheids" without 
attempting a finer discrimination;\\
\_ RR Lyrae-type variables: stars with periods between
0.2 and 1 day and light variations of about 1 magnitude. Among 
the known sub-types RRc stars
have periods between 0.2 and 0.5 days and amplitude of 0.5 magnitudes. 
Their
light curves have senoidal shapes. The RRab have periods between 0.4 and 
1 day,
with amplitudes up to 1 magnitude, and their light curves are 
asymmetric. Again,
we are not sensitive to shape details of the light curves and thus 
labeled these 
stars generically as "RR Lyrae".
\\

The result of this analysis can be appreciated in the tenth column of 
the Tables in
Appendix B, which constitutes our very preliminary classification of the 
variable stars
 and serve as a starting point for more comprehensive studies of these 
objects.


\section{Conclusions and future perspectives}

The development of a photometric survey with the Valinhos CCD Meridian 
Circle 
has permitted the identification of a significative number of 
previously unknown variables inside selected low-extinction windows 
towards the galactic bulge. Even if limited, the instrument could be 
used everyday and has a relatively simple
reduction data procedure, features that are clearly desirable for 
massive 
searches like ours. Other observational programmes with the aim of
detecting magnitude variations were also implemented  
like the monitoring of extragalactic sources brighter than 
$m_{Val}$ = 16 (Teixeira et al. 1998).

The use of Meridian Circle has other advantages: we were able to provide 
excellent positions for $\sim \, 30000$ stars 
(both variables and non-variables) among the 120000 
in the database, which are in most cases comparable to the
secondary catalog precisions (Table 2). This is a useful result 
for astrometric works, where a dense catalog of references is generally 
needed without any requirement of photometric stability. 
Proper motions researches, that are very important for the 
understanding of galactic kinematics are 
now possible and we pretend to contribute with more than 2000 frames 
of the windows
(including those without high photometric quality but 
good enough for astrometric purposes)  stored in CD ROM.

The {\it Class32} program, that was developed to organize the data and 
search for variable objects,
is of simple use and can be further improved. The main 
problem we found is the time of processing that made 
until now impossible a real-time analysis, although 
we are working on an updated version of the program for the 
implementation
of the project in a larger and more versatile instrument 
(likely a robotic 40 cm telescope). Some obvious modifications will be 
needed, for example: the first criterion of identification is 
strong and possible only because we were 
working with an astrometric instrument. The system of alarms and flags 
proved to be efficient and most of the spurious 
alerts were in fact caused by
the objects that are near the detection limit and had large photometric 
errors to be considered.

After all the reduction process and analysis, we found 
479 variable stars with light variations greater than 0.3 magnitudes
(the databases are continually
reprocessed and studied to complete and enlarge the catalogue).
Except for a small number of cases (like the Miras) the 
classification is tentative since we do not have spectral information 
as yet and our databases have small temporal coverage.
The final number of points on the light curve is smaller than originally 
expected and, in addition to the large photometric errors and the fact that 
an initial estimate of the range for period is unknown, its calculation 
was compromised.
In spite of these limitations, we were able to find period 
estimations for 79 variables (16.49 \% of the selected stars) using
the minimum entropy method (Cincotta et al. 1995).
We intend to use, in a future work, other methods (like PDM and modified 
Fourier method)
to optimize the results for our extreme conditions of large errors 
and few points.

In some light curves it was observed that the photometric errors are 
larger
for the brightest points (see the
light curve of the star LA2402 (Figure 13, Appendix A) for an example).
 We believe that this is caused by the effect of
chromatic aberration in the meridian circle, as briefly described in 
Appendix C.
The emission of these stars would be dominated by higher wavelengths and
the flux distribution would be wider and drops faster. This would 
provoke an 
increase in the flux error determination by the gaussian fit. 
Consequently, the magnitude errors should also increase.

 Thus, 97 variables were classified, among them, 21 Miras, 71 stars 
 classified as cepheids, 2 as RR Lyrae, 
 1 as an eclipsing binary and 3 as semi-regulars.
 Note that from our light curves, stars with periods smaller than 1 day 
and until 50
days, approximately, have the same shape because our temporal resolution 
and 
the few points obtained. Among the stars classified as "cepheids" many 
binaries 
probably occur. This should be the case of RR Lyrae too, but a more
accurate 
classification was not possible without spectral information. 
The star classified like eclipsing binary is already known from the 
GCVS.

We hope to refine this classification in future 
studies and contribute to other ongoing investigations involving these 
stars.

\begin{acknowledgements}

We would like to thank the financial support by FAPESP Foundation 
 (S\~ao Paulo, Brazil) through the grant 96/01477-8, and also CAPES, 
CNPq, Bordeaux Observatory, the Eletronics section of the IAG/USP,  
 P. Cincotta and R.D.D da Costa. 

\end{acknowledgements}

\appendix
\section{Light curves examples}

The following figures are representative of the light curves of the 479 
variable stars discovery and/or observed by the project. It is 
important to remember that since we worked with differential
magnitudes, the graphics do not display the apparent
magnitudes. Of course, the mean values of the magnitudes of
reference stars vary between nights. Table
A.1 can be used to have reference information about the apparent magnitude
of the stars ($m_{Val}$). The remaining light curves (not shown here), 
can be obtained electronically from the authors upon request.

\begin{table}[tb]
\caption{Mean magnitude of the reference stars for each window
that can be used to estimate the apparent magnitudes in light curves}
\label{tabrmm}
\begin{center}
\begin{tabular}{@{}lc@{}}
\hline
\rule{0pt}{1.2em}Window&  Mean magnitude ($m_{Val}$) \\
\hline
BE  &   11.888 \\  
BG  &   11.891 \\
BJ  &   11.804 \\
LA  &   11.619 \\
LB  &   11.626 \\
LC  &   11.513 \\
LD  &   11.591 \\
LI  &   11.961 \\
LR  &   11.612 \\
LT  &   11.627 \\
LU  &   11.685 \\
LV  &   12.052 \\
\hline
\end{tabular}
\end{center}
\end{table}

\bigskip

Figure 11: Light curves of the stars BG1159, LD610, BE1802 and LB1034. 

\bigskip

Figure 12: Light curves of the stars LD315, BE1152, BG1275 and LV95. 

\bigskip

Figure 13: Light curves of the stars LA2402, LA1925, BG801 and LR101.


\section{Valinhos Variable Stars Catalogues}

The Tables B.1 to B.12 contain the catalogue of variable stars discovered
and/or observed for the 12 low-extinction
fields towards the galactic bulge.

The first column of the catalog indicates  
the star label, formed by a name that identifies the window (see Table
1) and a sequential counter that indicates the
position of the object in the corresponding database. The following 
columns display successively 
the mean right ascension, the
mean declination, their standard deviations (in seconds and in 
arcseconds, respectively)(J2000);
 the mean magnitude observed ($m_{Val}$), the difference between the 
maximum and the
minimum magnitude value observed, the number of observations, a 
estimative for
the period, if possible, or an indication "NF" when the
stars show periodic characteristics but we are not able
to find a period, or "NC" when the star can be aperiodic or
have few observations.
The next column gives the tentative classification and the last are
the remarks about the previous known variables.

 Complete databases up to $m_{Val} \, = 16$ or better, with all
monitored stars 
will be available soon and will 
be published elsewhere. Finding charts will be available upon 
request from the authors.

\section{Distortion of the image profile by the chromatic aberration
in refractor instruments}

An interesting property of the Meridian Circle was noted and studied 
by us since the beginning of the programme in a parallel work. It is 
well-known that the image profile of refractor instruments, like the Valinhos 
Meridian
Circle, suffers distortions because of the objective chromatic aberration. This 
effect provokes that,
in the focal plane of the instrument (in which the CCD is placed), a 
{\it composition}
of several gaussians (which are the intersections of the flux 
distribution formed
at higher wavelengths) is actually observed. 
We display in Figure 14 an observed flux 
distribution and a tentative pure gaussian fit. As expected, the
gaussian does not fit the "wings" of the profile well since there is 
a non-negligible contribution from wavelengths other than the one 
for which focus is made.

Thus, the image profile reflects to 
a large extent intrinsic properties 
of the observed objects as modulated by the optical system. 
We are working to simulate the true (observed) profiles and
expect some correlation between the profile shape and spectral type. This 
can be a future tool to obtain quick 
spectral information in the single-filter 
observations of the Meridian Circle, 
limited to the brightest stars that are better sampled.

\bigskip

Figure 14: A tentative gaussian fit to the observed image
profile.


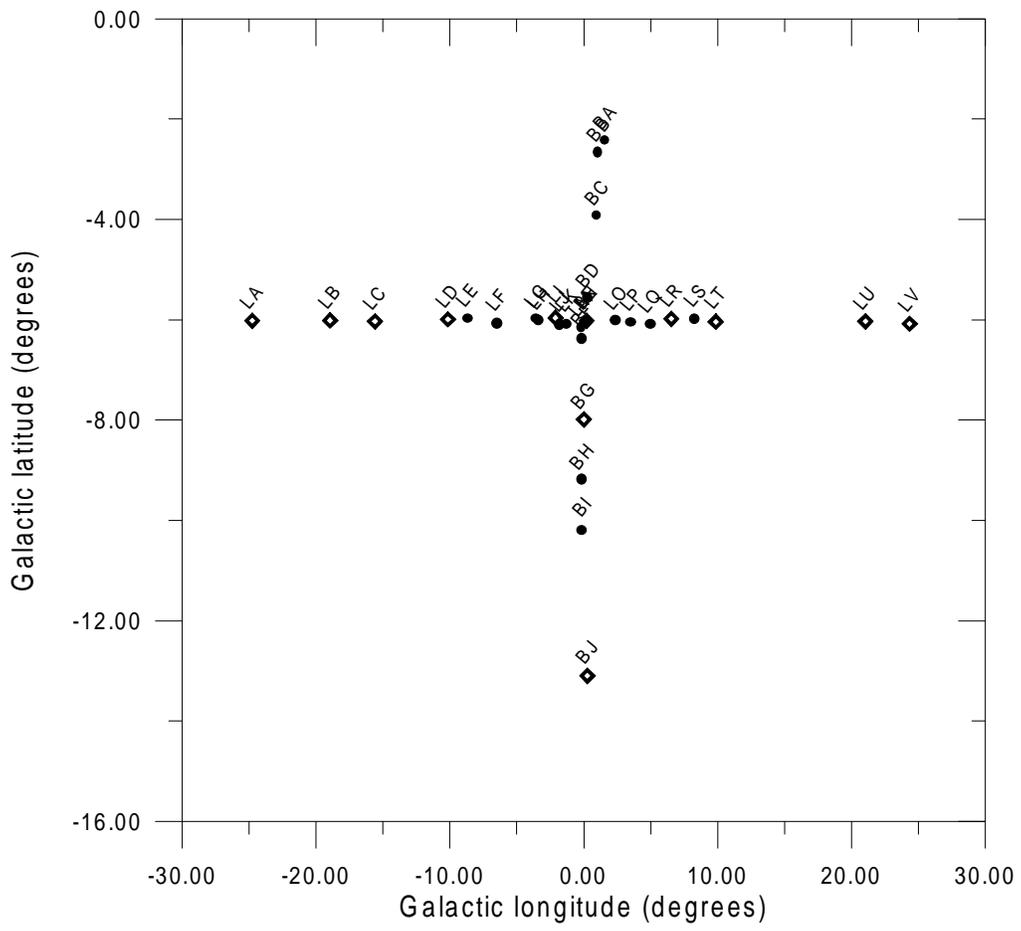
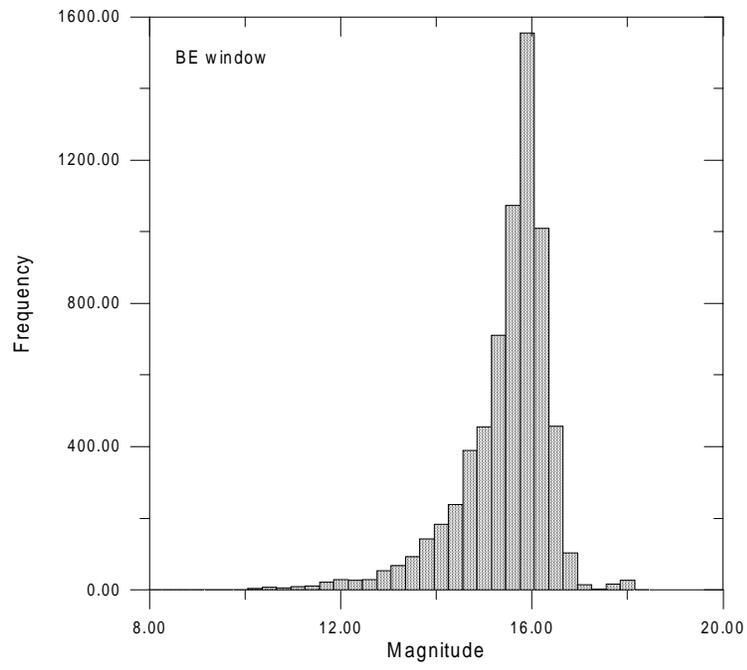
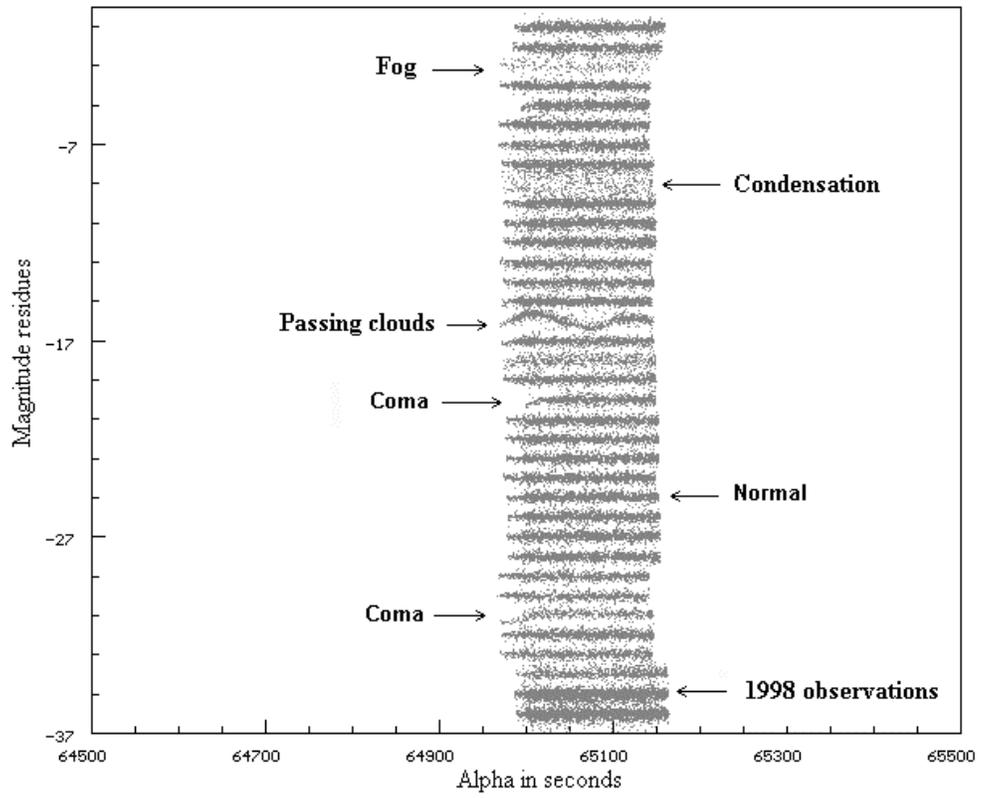
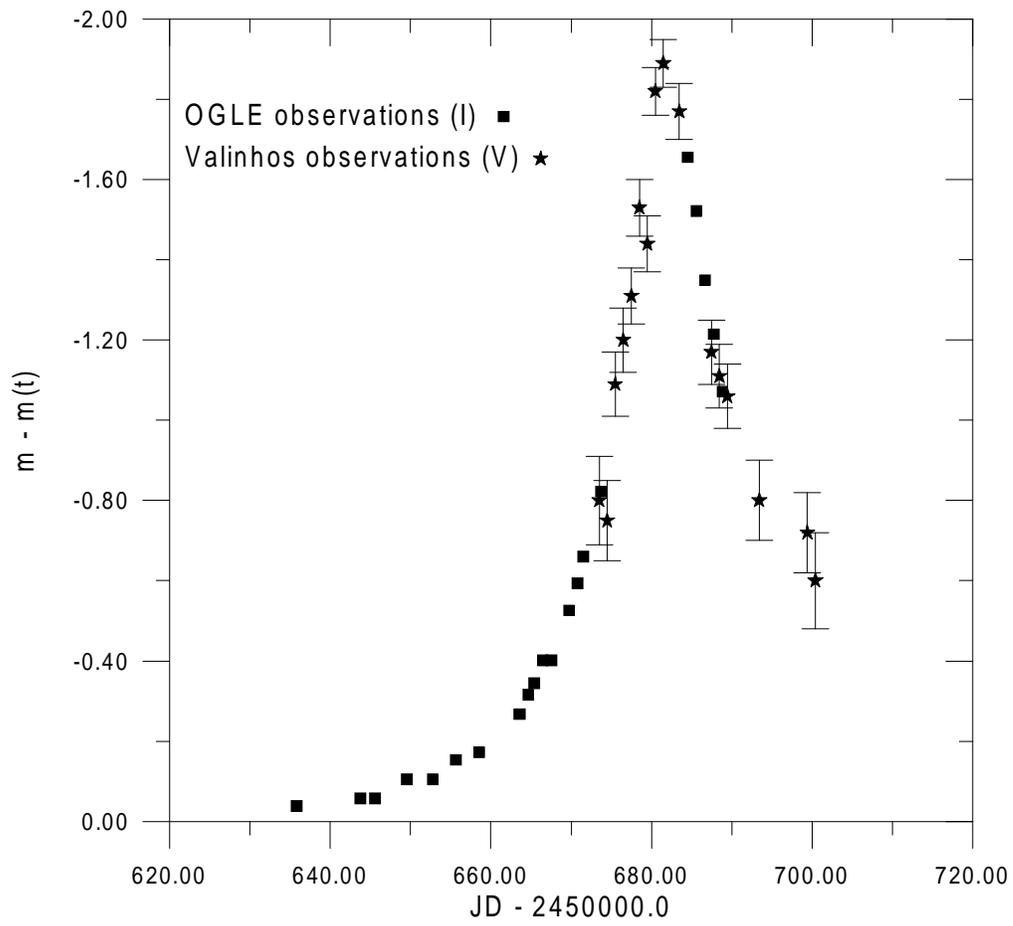
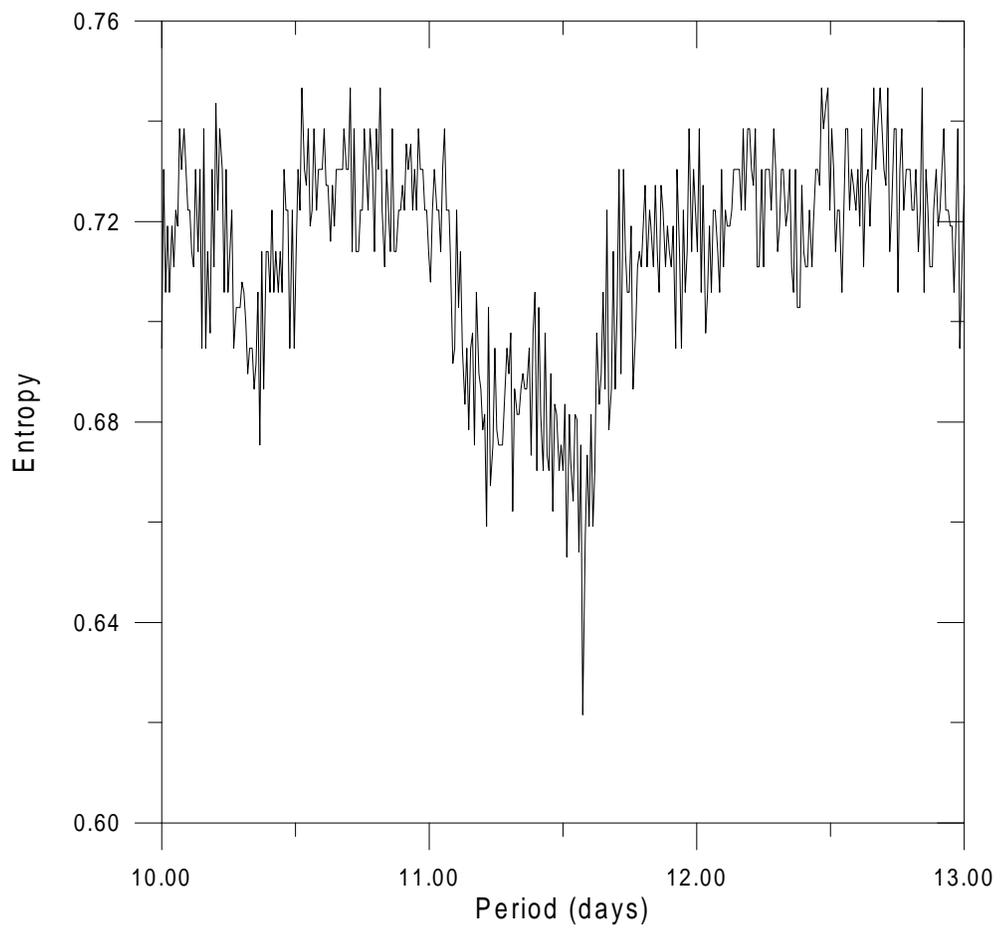
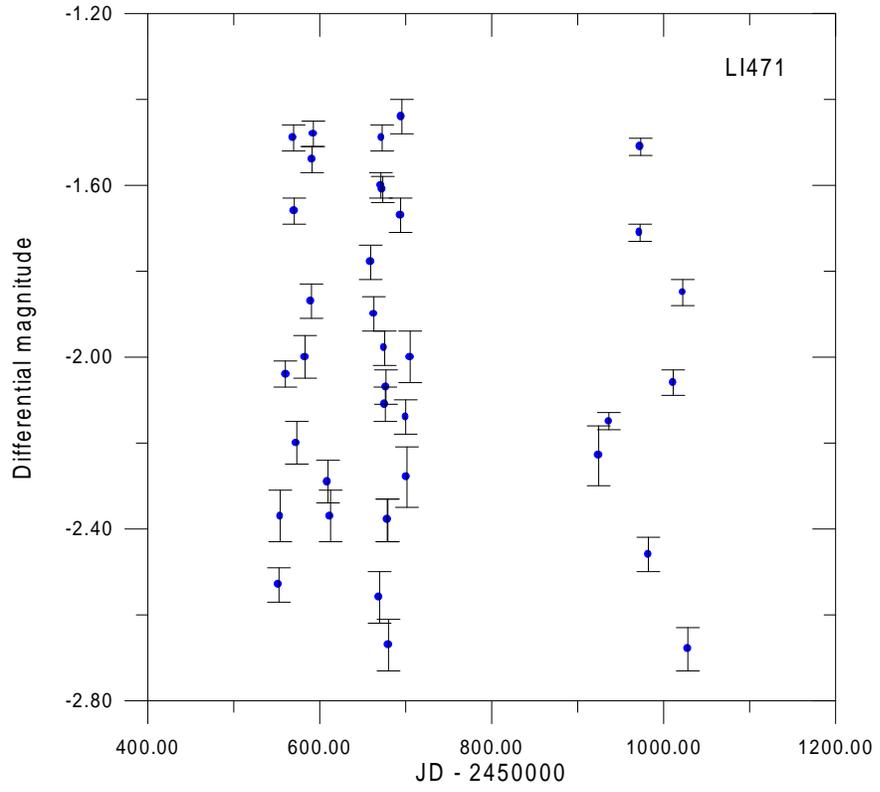
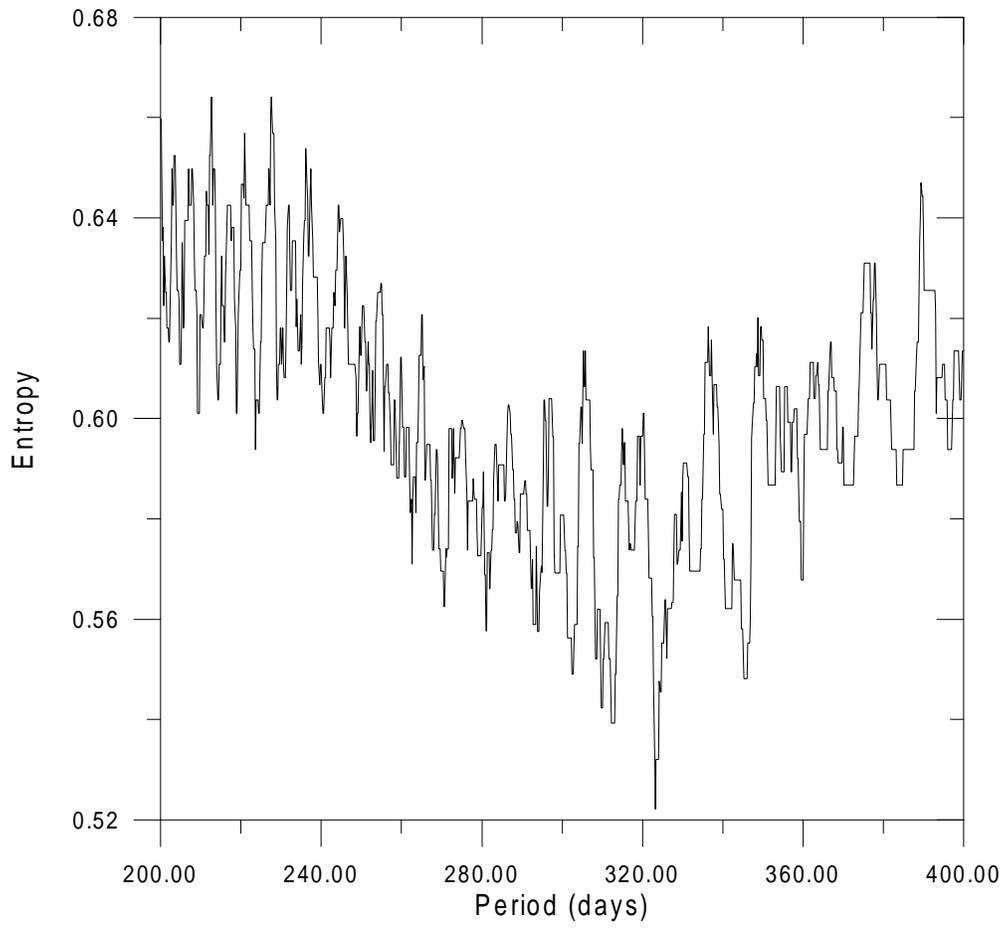
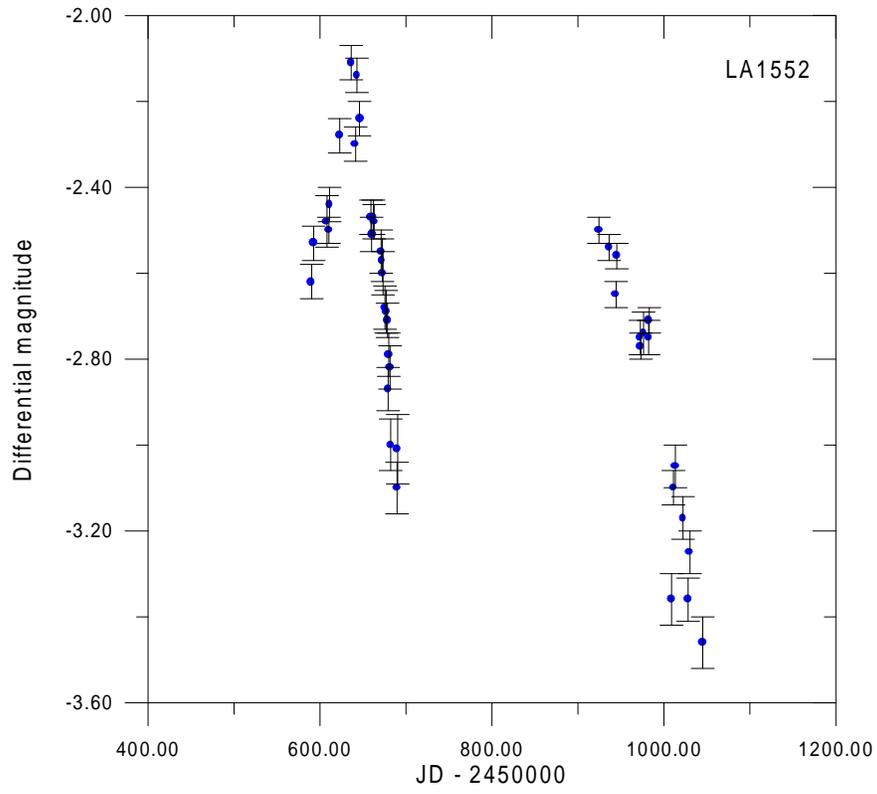
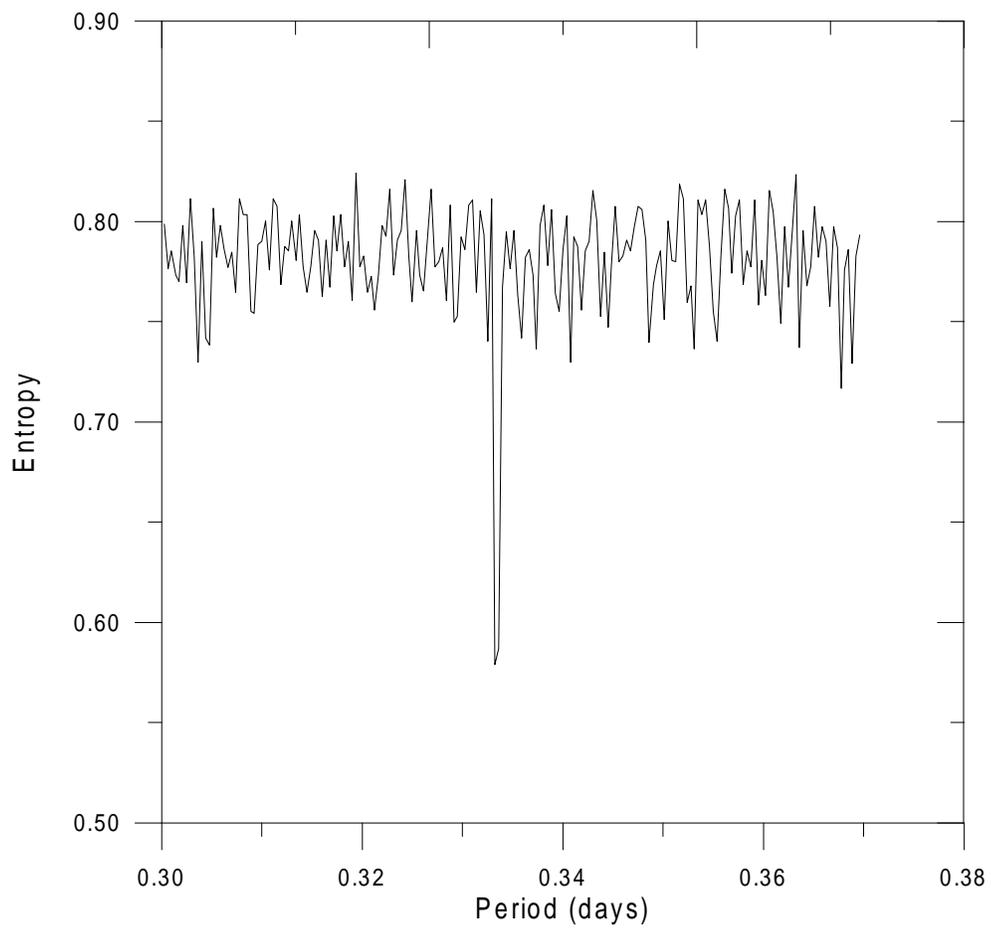
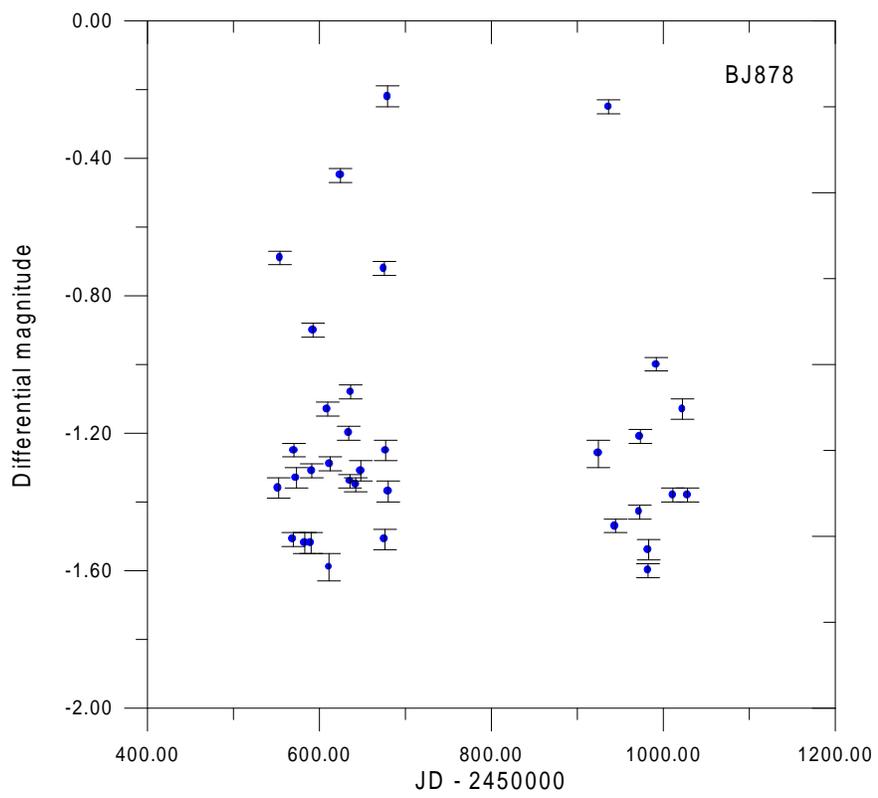
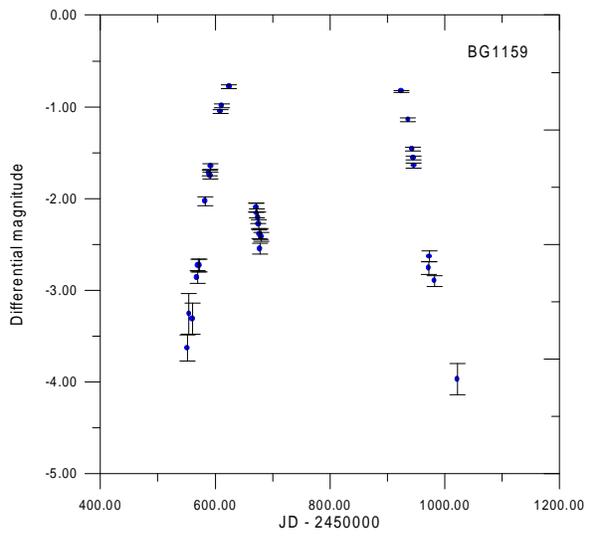
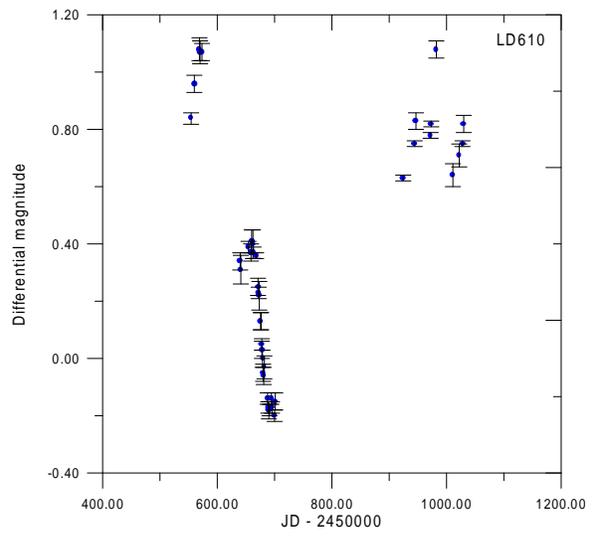
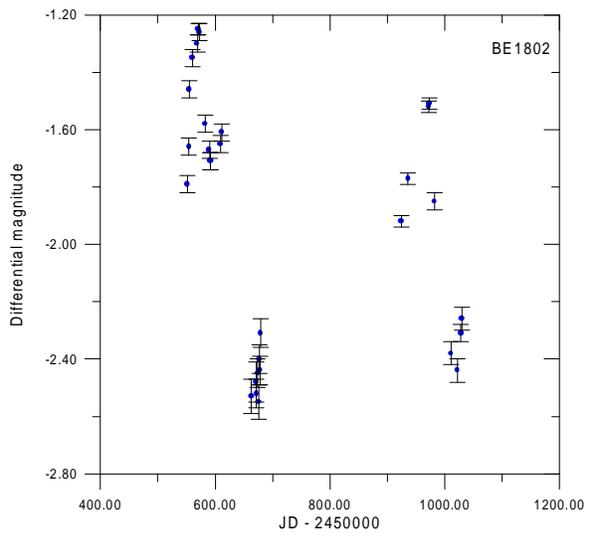
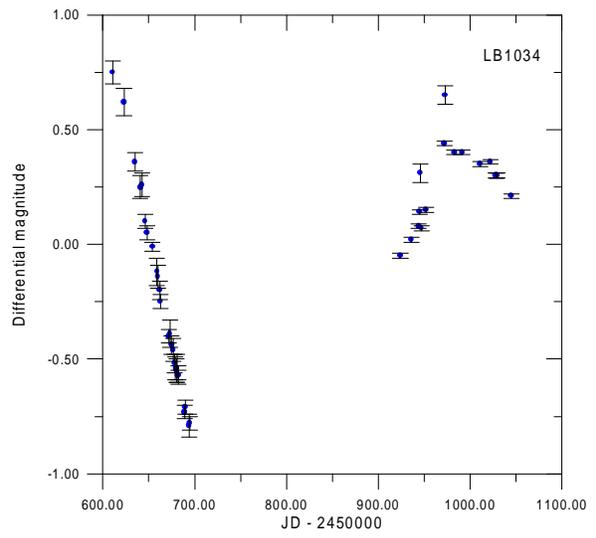
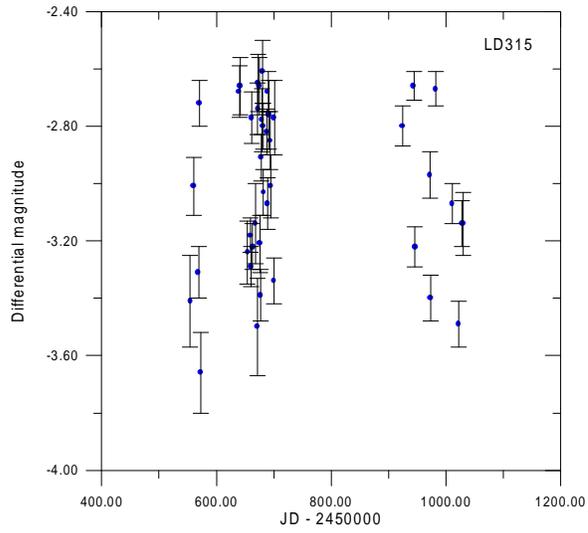
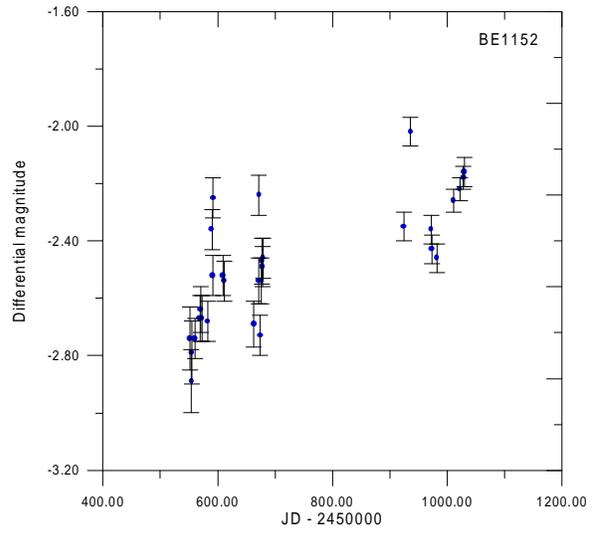
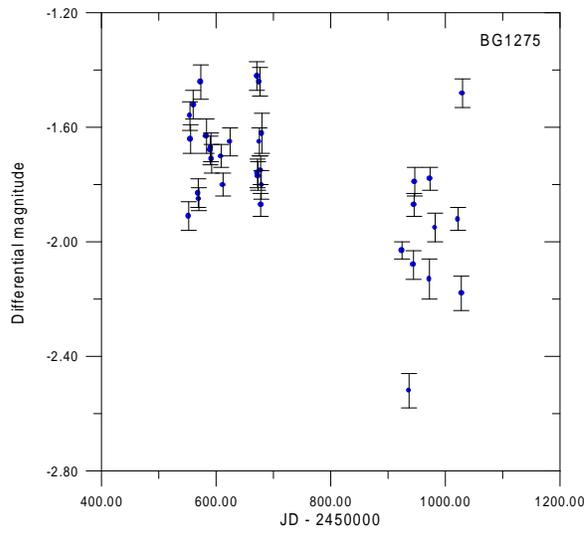
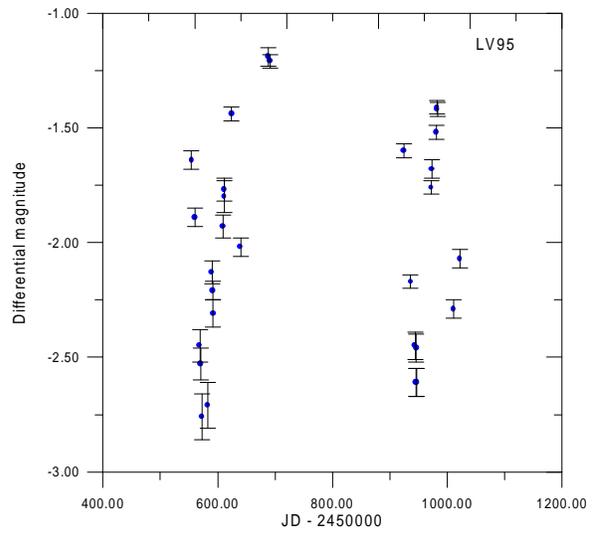
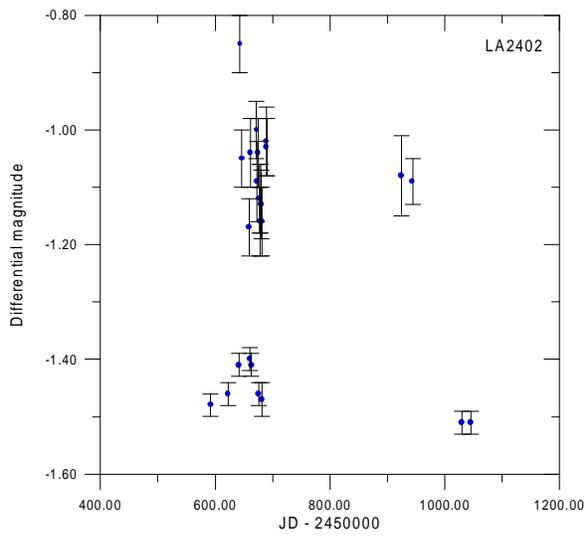
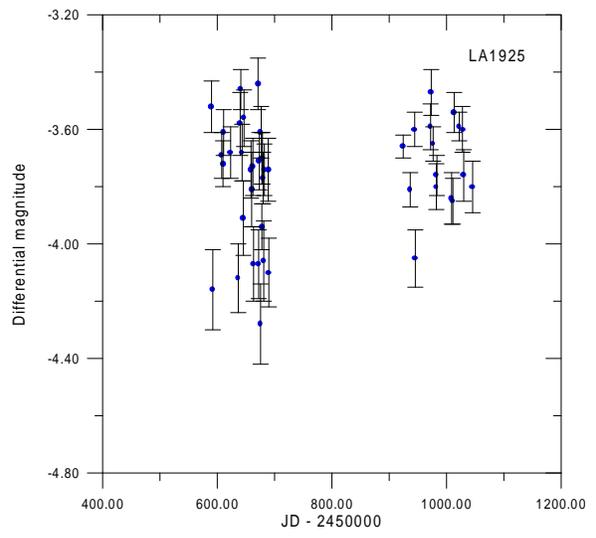
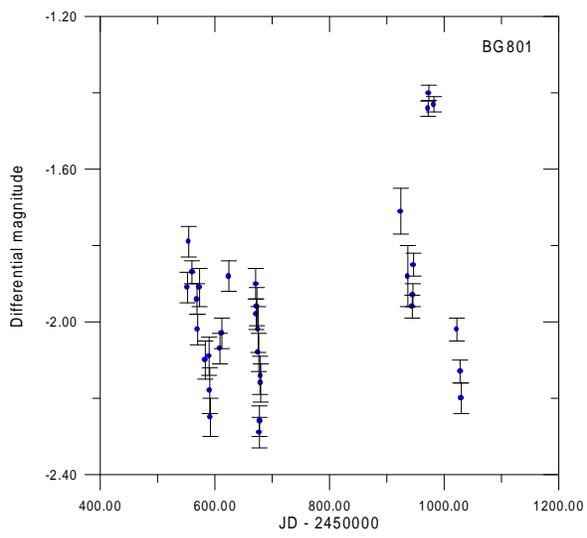
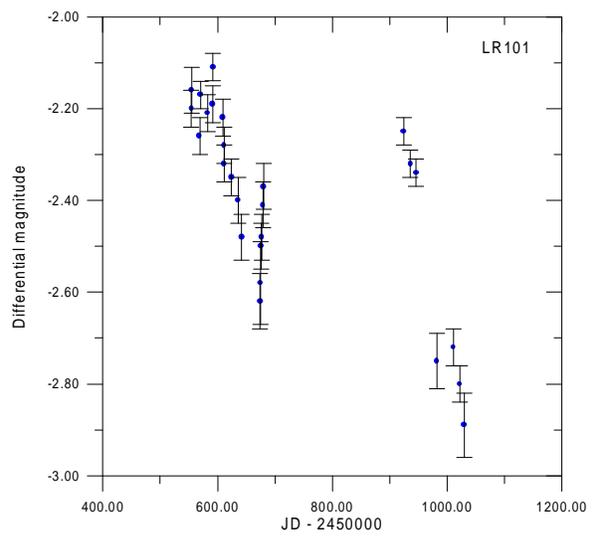
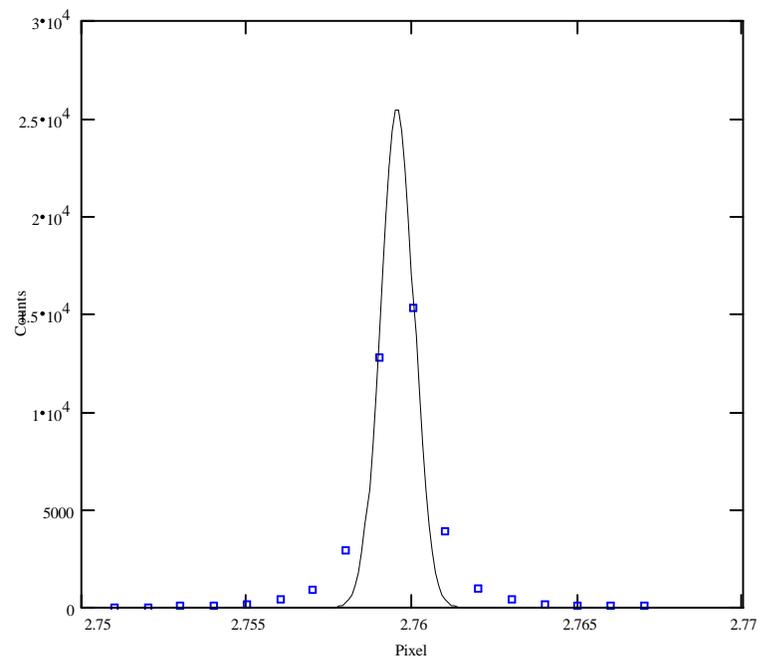
\end{document}